\documentclass[twocolumn, pra, superscriptaddress]{revtex4-1}
\usepackage{amssymb}
\usepackage{amsmath}
\usepackage{epsfig}
\usepackage{color}
\usepackage{graphics, graphicx}
\usepackage{bbold}
\usepackage{psfrag}
\usepackage{mathcomp}
\usepackage{subfigure}
\usepackage{verbatim}
\usepackage{color}
\usepackage[colorlinks,citecolor=blue]{hyperref}
\def\cp#1{\mathbf{#1}}

\begin{document}
\title{Fermion superfluid with hybridized $s$- and $p$-wave pairings}
\author{Lihong Zhou}
\affiliation{Beijing National Laboratory for Condensed Matter Physics, Institute of Physics, Chinese Academy of Sciences, Beijing 100190, China}
\author{Wei Yi}
\email{wyiz@ustc.edu.cn}
\affiliation{Key Laboratory of Quantum Information, University of Science and Technology of China,
CAS, Hefei, Anhui, 230026, People's Republic of China}
\affiliation{Synergetic Innovation Center of Quantum Information and Quantum Physics, University of Science and Technology of China, Hefei, Anhui 230026, China}
\author{Xiaoling Cui}
\email{xlcui@iphy.ac.cn}
\affiliation{Beijing National Laboratory for Condensed Matter Physics, Institute of Physics, Chinese Academy of Sciences, Beijing 100190, China}
\date{\today}

\begin{abstract}
Ever since the pioneering work of Bardeen, Cooper and Schrieffer in the 1950s, exploring novel pairing mechanisms for fermion superfluids has become one of the central tasks in modern physics. Here, we investigate a new type of fermion superfluid with hybridized $s$- and $p$-wave pairings in an ultracold spin-1/2 Fermi gas. Its occurrence is facilitated by the co-existence of comparable $s$- and $p$-wave interactions, which is realizable in a two-component $^{40}$K Fermi gas with close-by $s$- and $p$-wave Feshbach resonances. The hybridized superfluid state is stable over a considerable parameter region on the phase diagram, and can lead to intriguing patterns of spin densities and pairing fields in momentum space. In particular, it can induce a phase-locked $p$-wave pairing in the fermion species that has no $p$-wave interactions. The hybridized nature of this novel superfluid can also be confirmed by measuring the $s$-wave and $p$-wave contacts, which can be extracted from the high-momentum tail of the momentum distribution of each spin component. These results enrich our knowledge of pairing superfluidity in Fermi systems, and open the avenue for achieving novel fermion superfluids with multiple partial-wave scatterings in cold atomic gases.
\end{abstract}

\maketitle

\section{Introduction}
Fermion superfluid is one of the central research topics in modern physics. In recent years, ultracold Fermi gases have emerged as an excellent platform for the study of fermion superfluid in the strong-coupling regime via the Feshbach resonance (FR) technique~\cite{FR}.
Apart from the widely explored $s$-wave FRs, the $p$-wave FRs have also been realized in ultracold Fermi gases of $^{40}$K~\cite{K40,Toronto} or $^{6}$Li~\cite{Li6} atoms. The $s$- and $p$-wave FRs are associated with distinct fermion superfluids. Across an $s$-wave FR, the pairing superfluid undergoes a smooth crossover from the Bardeen-Cooper-Schrieffer (BCS) regime with weakly bound Cooper pairs to the Bose-Einstein condensation regime with tightly bound molecules~\cite{Leggett, review}. While across a $p$-wave FR, the pairing superfluid can go through a phase transition which is characterized by a change of the pairing field orientation with respect to the external magnetic field~\cite{Gurarie, Yip}. Such a dramatic difference originates from their contrastive pairing symmetries: for the $s$-wave case, a Cooper pair is a spin singlet with isotropic orbitals regardless of the interaction strength; the $p$-wave case, however, features spin-triplet pairing with anisotropic orbitals, which renders the pairing very sensitive to the relative interaction strengths along different orbital orientations.

\begin{figure}[t]
\includegraphics[width=8cm]{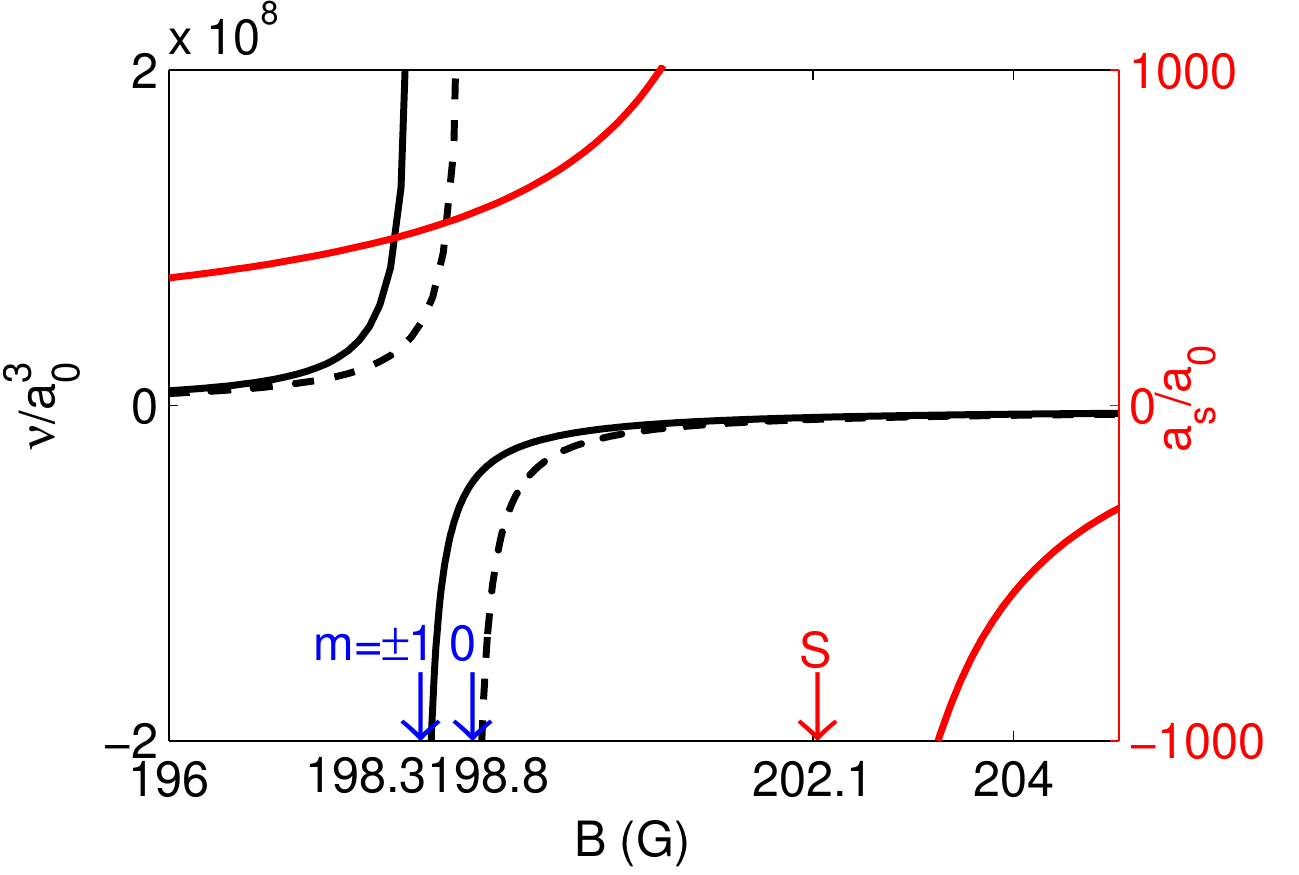}
\caption{(Color online). Feshbach resonances(FR) of $^{40}$K atoms. Black lines show the $p$-wave scattering volume $\nu$ (in unit of $a_0^3$, $a_0$ is Bohr radius) and red line shows the $s$-wave scattering length $a_s$ (in unit of $a_0$).  The $s$-wave FR between atomic hyperfine states $|F=9/2,m_F=-7/2\rangle\equiv|\uparrow\rangle$ and $|F=9/2,m_F=-9/2\rangle\equiv|\downarrow\rangle$ occurs at $B=202.1$G (with a width of $8$G), which is very close to the $p$-wave FRs between two $|\uparrow\rangle$s at $198.3$G and $198.8$G (with a width of $0.5$G), respectively with orbital angular momentum $l=1,m=\pm 1$ and $l=1,m=0$.
} \label{fig1}
\end{figure}

Near a $p$-wave FR, atom losses have been generally considered to
prohibit a global equilibration throughout the system. However, a
quasi-equilibration can still be reached at short time scales, as
has been demonstrated in a Bose-Einstein condensate following a
quench into regimes with large $s$-wave scattering length and
considerable atom losses~\cite{JILABECWORK}. Indeed, in a very
recent experiment~\cite{Toronto}, a steady state with strong
$p$-wave correlations has been achieved in a Fermi gas at a low
temperature $T=0.2 T_F$ ($T_F$ is the Fermi temperature) within a
time scale of $\sim 0.5$ms. It is thus hopeful that pairing physics
can be explored near a $p$-wave FR at lower temperatures, where a
steady state with interesting pairing correlations can be probed
before atom losses become dominant.

In this work, we investigate a new type of fermion superfluid, where
the aforementioned two distinct pairing symmetries co-exist and
hybridize with each other. Such a superfluid is facilitated by the
presence of both $s$- and $p$-wave interactions with comparable
strengths, which can be realized in Fermi gases of $^{40}$K atoms
near a magnetic field of $B\approx 198$G~\cite{K40}. As illustrated
in Fig.~\ref{fig1}, for the two hyperfine states
$|F=9/2,m_F=-7/2\rangle\equiv|\uparrow\rangle$ and
$|9/2,-9/2\rangle\equiv|\downarrow\rangle$, the $s$-wave (between
$|\uparrow\rangle$ and $|\downarrow\rangle$) and the $p$-wave
(between two $|\uparrow\rangle$s) FRs are sufficiently close to each
other. This physical system offers a promising platform to
investigate the effects of multiple partial-wave scatterings on the
pairing superfluidity of fermions, as we will address in this work.

To characterize the novel hybridized fermion superfluid, we first consider the case of isotropic $p$-wave interactions. We show that in the hybridized superfluid state, the spin densities and the pairing fields exhibit intriguing patterns in momentum space that are drastically different from those of a conventional superfluid. We find an induced $p$-wave pairing between the $|\downarrow\rangle$ states, which have no $p$-wave interactions. Interestingly, the phases of the $p$-wave pairing between the $|\uparrow\rangle$ states and that between the $|\downarrow\rangle$ states are locked to be conjugate with each other through the hybridization with the $s$-wave pairing. Such a hybridization can also manifest itself in the measurement of the $s$-wave and $p$-wave contacts, which can be extracted from the high-momentum tail of the momentum distribution of each spin component. Finally, we point out the rich orbital structures of the hybridized superfluid states under typical experimental conditions with anisotropic $p$-wave interactions.

\section{Formalism}
We start from a two-channel Hamiltonian, $\Omega=H-\sum_{\sigma} \mu_{\sigma} N_{\sigma}$, for spin-1/2 fermions:
\begin{align}
\Omega=&\sum_{\cp{k}, \sigma=\uparrow,\downarrow}(\varepsilon_{\cp k }-\mu_{\sigma})a_{\cp k \sigma}^\dagger a_{\cp k \sigma}+\sum_{\cp{q},\alpha}(\varepsilon_{\cp q}^b+\epsilon_{\alpha}-2\mu_{\uparrow})b_{\cp {q}\alpha}^\dagger b_{\cp {q}\alpha}\nonumber\\
&+\frac{U}{V}\sum_{\cp{kk'p}} a_{\cp {k},\uparrow}^\dagger a_{\cp {p-k},\downarrow}^\dagger a_{\cp {p-k'},\downarrow}a_{\cp{k}',\uparrow}
\nonumber\\
&+\sum_{\cp{kq}\alpha}\frac{g(|{\cp k}|)}{\sqrt{V}} \cp{k}_\alpha(b_{\cp {q}\alpha}^\dagger a_{\frac{\cp {q}}{2}+\cp{k},\uparrow}a_{\frac{\cp {q}}{2}-\cp{k},\uparrow}+h.c.), \label{hamiltonian}
\end{align}
with $N_{\downarrow}=\sum_{\cp k} a_{\cp k \downarrow}^\dagger a_{\cp k \downarrow},\ N_{\uparrow}=\sum_{\cp k} a_{\cp k \uparrow}^\dagger a_{\cp k \uparrow}+2\sum_{{\cp q},\alpha}b_{\cp {q}\alpha}^\dagger b_{\cp {q}\alpha} $. Here, $a_{\cp{k},\sigma}^\dagger$ is the creation operator of spin-$\sigma$ atom with momentum ${\cp k}$ and energy $\varepsilon_{\cp k }={\cp k}^2/(2M)$;  $b_{\cp {q}\alpha}^\dagger$ is the creation operator of a $p$-wave bosonic molecule with momentum ${\cp q}$, kinetic energy $\varepsilon_{\cp q }^b={\cp q}^2/(4M)$, and detuning $\epsilon_{\alpha}$ (the direction of spin-polarization $\alpha=x,y,z$); $g(|{\cp k}|)=g\theta(\Lambda-|{\cp k}|)$ is the $p$-wave coupling between a bosonic molecule and two spin-$\uparrow$ atoms with relative momentum ${\cp k}$ (here $\Lambda$ is the momentum cutoff, $\theta(x)$ is the Heaviside step function). The bare $p$-wave coupling $g$, the detuning $\epsilon_{\alpha}$, and the momentum cutoff $\Lambda$ are related to the scattering volume $\nu_{\alpha}$ and the effective range $k_0$~\cite{Gurarie2}:
\begin{align}
\frac{1}{\nu_{\alpha}} &=-\frac{6\pi \epsilon_{\alpha}}{M g^2}  + \frac{2}{3\pi} \Lambda^3 , \\
k_0 &=-\frac{12\pi}{M^2 g^2}  - \frac{4}{\pi} \Lambda.
\end{align}
$U$ gives the bare $s$-wave interaction, which is related to the $s$-wave scattering length $a_s$ by the renormalization relation:  $1/U=M/(4\pi a_s)-1/V \sum_{\cp k} 1/(2\varepsilon_{\cp k })$, with $V$ the volume of the system. In this work we set $\hbar=1$ for convenience.

Based on the standard BCS theory, we define two pairing fields $\Delta_{\cp{p}}=\frac{U}{V}\sum_{\cp{k}}\langle a_{\cp{p-k},\downarrow}a_{\cp{k},\uparrow}\rangle$ and $\lambda_{\cp{q}\alpha}=\frac{g}{\sqrt{V}} \langle b_{\cp{q}\alpha}\rangle$, which are respectively the pairing order parameters of the $s$- and the $p$-wave superfluids. In this work we consider zero total momentum for each pairing state~\cite{note_zeromomenta}, and denote $\Delta_{\cp{p}=0}\equiv \Delta$, $\lambda_{\cp{q}=0,\alpha}\equiv \lambda_{\alpha}$. The Hamiltonian can then be written as $\Omega=\sum_{\cp{k}>0}\psi_{\cp{k}}^\dagger H_{\cp{k}}\psi_{\cp{k}}+{\rm Const.}$, where
the vector operator $\psi_{\cp{k}}=\left(a_{\cp{k},\uparrow}, a_{-\cp{k},\uparrow}^\dagger , a_{\cp{k},\downarrow} , a_{\cp{-k},\downarrow}^\dagger\right)^T$, and the matrix $H_{\cp k}$ is given by:
  \begin{align}
\left(
  \begin{array}{cccc}
    \varepsilon_{\cp k}-\mu_{\uparrow}& -2\theta(\Lambda-|{\cp k}|)\bar{\lambda}&0&\Delta\\
    -2\theta(\Lambda-|{\cp k}|)\bar{\lambda}^\ast& -\varepsilon_{\cp {-k}}+\mu_{\uparrow}&-\Delta^\ast&0\\
    0&-\Delta&\varepsilon_{\cp {k}}-\mu_{\downarrow}&0\\
    \Delta^\ast&0&0&-\varepsilon_{\cp {-k}}+\mu_{\downarrow}\\
  \end{array}
\right)  \nonumber
\end{align}
with $\bar{\lambda}=\sum_{\alpha}k_\alpha\lambda_{\alpha} $. By diagonalizing the matrix as $\bar{H}_{\cp k}= S^{-} H_{\cp k} S\equiv {\rm Diag}(E_{{\cp k}1}, E_{{\cp k}2}, E_{{\cp k}3}, E_{{\cp k}4})$ with the eigen operator $\bar{\psi}_{\cp{k}}=  S^- \psi_{\cp{k}}\equiv \left( \alpha_{\cp{k}}, \alpha_{-\cp{k}}^\dagger, \beta_{\cp{k}}, \beta_{\cp{-k}}^\dagger \right)^{\rm T} $, $\Omega$ can be reduced to:
 \begin{align}
\Omega=&\sum_{\cp{k}>0}\left[ \sum_{i=1}^4 E_{{\cp k}i}\theta(-E_{{\cp k}i}) + 2(\varepsilon_{\cp {k}} -\mu)\right] \nonumber\\
&\ \ \ +\sum_{\alpha}\frac{ \epsilon_{\alpha}-2\mu-2h}{g^2} |\lambda_{\alpha}|^2-\frac{V}{U}|\Delta|^2,
 \end{align}
where we have used $\mu=(\mu_{\uparrow}+\mu_{\downarrow})/2,\  h=(\mu_{\uparrow}-\mu_{\downarrow})/2$. The ground state of the system can then be determined by minimizing $\Omega$ in terms of $\Delta$ and $\lambda_{\alpha}$. Due to the gaugeable global phase for both the $s$- and the $p$-wave pairing fields, we set $\Delta$ to be real and denote the vector $\vec{\lambda}\equiv(\lambda_x,\lambda_y,\lambda_z)$ as $\vec{\lambda}=\vec{u}+i\vec{v}$ with $\vec{u}\cdot\vec{v}=0$~\cite{Gurarie}.

\section{Phase diagram}
To capture the essential physics of a hybridized superfluid, we first consider the simple case of isotropic $p$-wave interactions, i.e., all $\epsilon_{\alpha}$ are equal. It follows that $\vec{u}$ and $\vec{v}$ have a simultaneous SO(3) rotational symmetry in the coordinate space. For convenience, we choose $\vec{u}=u \hat{z}$ and $\vec{v}=v\hat{x}$. In this case, the $p$-wave superfluid is always associated with the $p+ip$ pairing symmetry, reminiscent of the $A$ phase in $^{3}$He~\cite{Anderson,Ho}.

\begin{figure}[h]
\includegraphics[width=8.5cm]{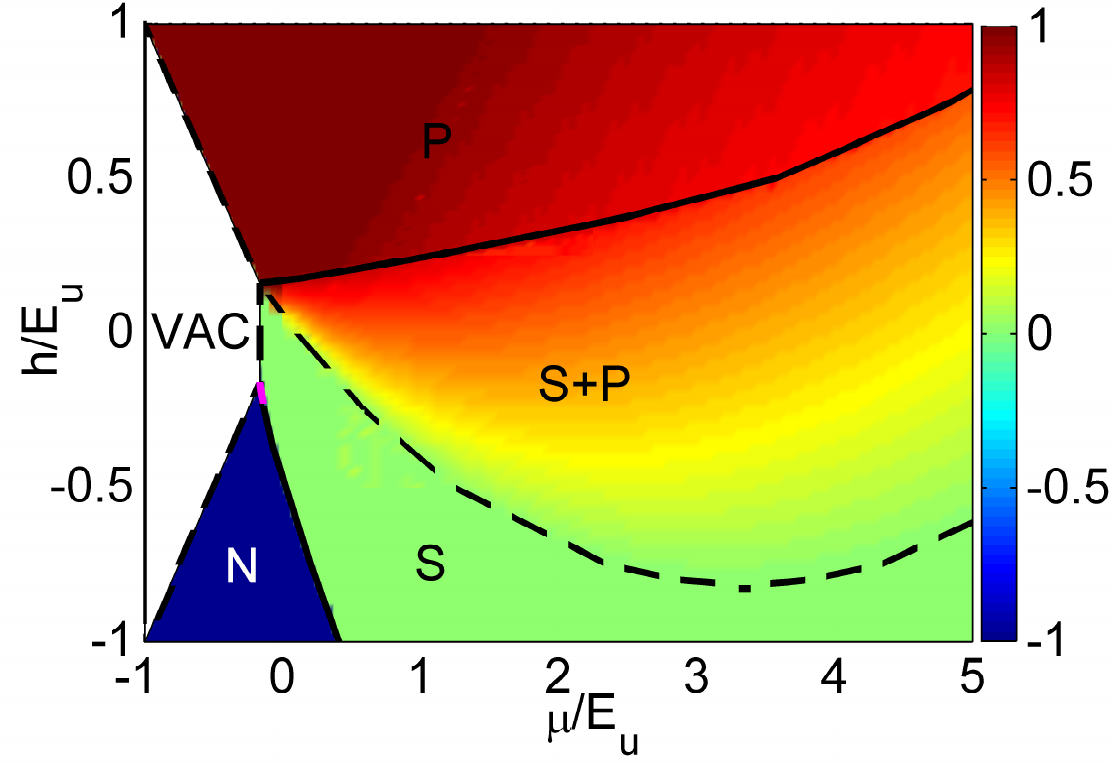}
\caption{(Color online). (a) Ground-state phase diagram in the ($\mu,h$) plane with interaction parameters $a_s=25/|k_0|$, $1/\nu_{\alpha}=0$,  and the $p$-wave cutoff $\Lambda=0.75 |k_0|$. $\mu, h$ are in the unit of $E_u=k_u^2/(2M)$, with $k_u=|k_0|/10$. The diagram includes the vacuum (VAC), the normal phase (N) and various superfluid phases (S, P, S+P). The solid (dashed) lines are the phase boundaries for first-order (continuous) transitions. The magenta region in-between N and S is the breached pairing phase~\cite{bp1,bp2,Bedaque,Sheehy, Hu,Pao,bp4}, which is only stabilized in a small region in the phase diagram with $\mu/E_{2b}\in (-0.5, -0.4)$ (here $E_{2b}=1/(M a_s^2)$ is the two-body binding energy). The background color denotes the spin polarization $P=(n_{\uparrow}-n_{\downarrow})/(n_{\uparrow}+n_{\downarrow})$.
}
\label{fig2}
\end{figure}

In Fig.~\ref{fig2}, we show a typical ground-state phase diagram in terms of the chemical potentials $(\mu,h)$ at the $p$-wave resonance $1/\nu_{\alpha}=0$. Based on the parameters of $^{40}$K atoms near the $B\approx 198$G $p$-wave resonance with $k_0=-8\times 10^8$m$^{-1}$ and $a_s =3.15\times 10^{-8}$m ~\cite{K40}, we set the interaction parameter $|k_0|a_s= 25$ and the p-wave cutoff $\Lambda=0.75|k_0|$, and use $k_u=k_0/10$ as the momentum unit. We can see that by adjusting $\mu$ and $h$, the system can exhibit various superfluid phases, including the purely $s$-wave superfluid with $\Delta\neq 0, \ u=v=0$ (S), and the purely $p+ip$ superfluid with $\Delta= 0, \ u=v\neq0$ (P). Of particular interest here is the superfluid with co-existing $s$- and $p$-wave pairing orders: $\Delta\neq0, \ u=v\neq0$ (S+P), which emerges in-between S and P phases on the phase diagram. By examining the continuity property of the pairing amplitudes ($\Delta, u,v$) and the spin densities across the phase boundaries, we find that the transition between S and S+P is continuous, while that between S+P and P is of first order. These phase transitions can be observed directly by measuring density profiles of a trapped gas. As suggested by Fig.~\ref{fig2}, from the trap center to the edge there could exist different sequences of phases, for instance, (S+P)--P (for $N_{\uparrow}>N_{\downarrow}$) or (S+P)--S--Normal (for $N_{\uparrow}<N_{\downarrow}$). At the (S+P)--P phase boundary the spin densities change discontinuously, while at the (S+P)--S boundary the spin polarization continuously goes to zero.

\section{Superfluid with hybridized pairings}
In the following, we focus on the exotic S+P superfluid state, where both the $s$- and $p$-wave pairings are present. As shown in Fig.~\ref{fig2}, this state exists for a spin-imbalanced system with more spin-$\uparrow$ than spin-$\downarrow$ atoms. However, the characterization of this state is far beyond a simple superposition of the $s$-wave pairing between an equal number of different spins and the $p$-wave pairing in the remaining spin-$\uparrow$ atoms. We will demonstrate that the two pairing fields are in fact coherently entangled with each other, which leads to dramatic physical consequences as shown below.

We start from the ground state wave function:
\begin{equation}
|\Psi_G\rangle=\prod_{{\cp k}>0} \prod_{i=1}^4 \gamma_{{\cp k},i} |{\rm vac}\rangle ,
\end{equation}
with $\gamma_{{\cp k},1}=\theta(E_{\cp{k}1}) \alpha_{\cp k}+\theta(-E_{\cp{k}1}) \alpha_{\cp k}^{\dag}$, $\gamma_{{\cp k},2}=\theta(-E_{\cp{k}2}) \alpha_{-\cp {k}}+\theta(E_{\cp{k}2}) \alpha_{-\cp k}^{\dag}$, $\gamma_{{\cp k},3}=\theta(E_{\cp{k}3}) \beta_{\cp k}+\theta(-E_{\cp{k}3}) \beta_{\cp k}^{\dag}$, $\gamma_{{\cp k},4}=\theta(-E_{\cp{k}4}) \beta_{-\cp k}+\theta(E_{\cp{k}4}) \beta_{-\cp k}^{\dag}$. For any given ${\cp k}$ and $E_{{\cp k}i}$, one can expand $\gamma_{{\cp k},i}$ in terms of $\{a_{\cp{k}\sigma}, a_{\cp{k}\sigma}^{\dag}\}$, and finally we arrive at a general form of the wave function:
\begin{align}
&|\Psi_G\rangle=\prod_{{\cp k}>0}  \Big( u_{\cp k}+v^{\uparrow\uparrow}_{\cp{k}} a_{\cp{k},\uparrow}^{\dag} a_{\cp{-k},\uparrow}^{\dag} + v^{\downarrow\downarrow}_{\cp{k}} a_{\cp{k},\downarrow}^{\dag} a_{\cp{-k},\downarrow}^{\dag} +\nonumber\\
 & v^{\uparrow\downarrow}_{\cp{k}} a_{\cp{k},\uparrow}^{\dag} a_{\cp{-k},\downarrow}^{\dag} +
 v^{\downarrow\uparrow}_{\cp{k}} a_{\cp{k},\downarrow}^{\dag} a_{\cp{-k},\uparrow}^{\dag}
 +  v^{(4)}_{\cp k}a_{\cp{k},\uparrow}^{\dag} a_{\cp{-k},\downarrow}^{\dag} a_{\cp{-k},\uparrow}^{\dag} a_{\cp{k},\downarrow}^{\dag}  \Big) |{\rm vac}\rangle .
\end{align}
It can be easily checked that for a pure $s$- or $p$-wave superfluid, the intra- and inter-species pairing terms, $v_{\cp k}^{\sigma \sigma}$ and $v_{\cp k}^{\sigma \bar{\sigma}}$, cannot exist simultaneously for any given ${\cp k}$; while for the hybridized superfluid they can co-exist and produce non-zero order parameters for both the $s$- and $p$-wave pairings. Furthermore, the co-existence of $u_k$ and $v_{\cp k}^{\downarrow\downarrow}$ implies an induced $p$-wave pairing within the spin-$\downarrow$ atoms, which, according to the original Hamiltonian (Eq.~\ref{hamiltonian}), do not have any $p$-wave interactions at all.

\begin{widetext}
\begin{figure*}[t]
\includegraphics[width=17cm]{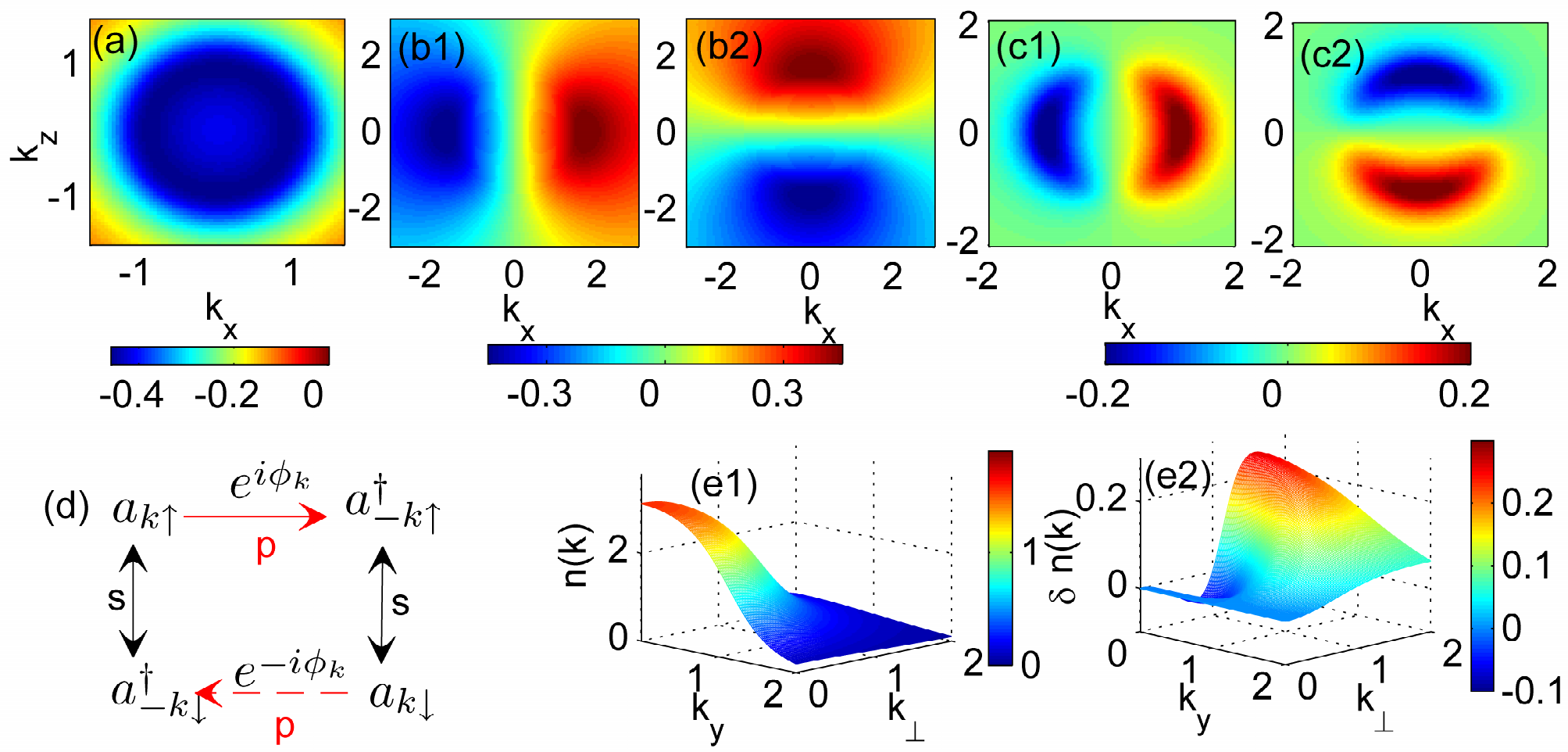}
\caption{(Color online). Contour plots of pairing fields and density distributions in the momentum space ($k_y=0$) for the hybridized superfluid phase at $\mu=E_u$ and $h=0$. (a) is for $s$-wave pairing; (b1) and (b2) are the real and imaginary parts of $p$-wave pairing between spin-$\uparrow$; (c1) and (c2) are the real and imaginary parts of the induced $p$-wave pairing between spin-$\downarrow$. (d) Schematics for the mechanism of induced $p$-wave pairing with a locked phase. (e1, e2) momentum distributions of total density $n(\cp k)$ and spin density $\delta n(\cp k)$. All momenta are in the unit of $k_u$, and all densities are in unit of $1/k_u^3$.}
\label{fig3}
\end{figure*}
\end{widetext}

In Fig.~\ref{fig3}(a-c), we show the typical contour plots of $s$- and $p$-wave pairing fields in momentum space for the hybridized superfluid phase. We can see that the $s$-wave pairing $\langle a_{-\cp{k},\downarrow} a_{\cp{k},\uparrow} \rangle$ is always isotropic in momentum space (Fig.~\ref{fig3}(a)), while the $p$-wave pairing $\langle a_{-\cp{k},\uparrow} a_{\cp{k},\uparrow} \rangle$ (b1,b2) and the induced $p$-wave pairing $\langle a_{-\cp{k},\downarrow} a_{\cp{k},\downarrow} \rangle$ (c1,c2) show quite novel structures compared to the purely $p$-wave case. In particular, the phases of these two $p$-wave pairing fields are locked to be conjugate to each other, i.e., if the pairing between $\uparrow,\uparrow$ is of the type of $p+ip$, then the pairing between $\downarrow,\downarrow$ will be $p-ip$. This can be seen clearly from the real and imaginary parts of pairing fields in (b1,b2) and (c1,c2).

The correlated $p$-wave pairings in different spin species is an essential consequence of the hybridization of $s$- and $p$-wave pairings. As schematically shown in Fig.~\ref{fig3}(d), a $p+ip$ superfluid quasi-particle is superposed by $a_{\cp{k},\uparrow}$ and $a_{-\cp{k},\uparrow}^{\dag}$ with a phase $e^{i\phi_{\cp k}} (\phi_{\cp k}=\arctan(k_z/k_x))$ inbetween; when they individually couple to $a_{-\cp{k},\downarrow}^{\dag}$ and $a_{\cp{k},\downarrow}$ by $s$-wave pairing, the phase is effectively transferred to these spin-$\downarrow$ operators, and finally a conjugate phase ($e^{-i\phi_{\cp k}}$) is produced, which gives rises to the induced $p-ip$ pairing between spin-$\downarrow$ atoms. The mutual phase locking between the two $p$-wave pairing fields is intrinsic to the $s$- and $p$-wave hybridized pairings, regardless of the symmetry of the $p$-wave interaction (isotropic or anisotropic in orbitals).

We note that the intriguing phenomenon of induced pairing is a unique feature of our system. It is fundamentally different from the co-existence of pairing correlations reported in previous studies. For example, the co-existence of $s$- and $p$-wave pairings has been investigated in a fermion superfluid with long-range dipole-dipole interactions, which naturally contain all partial-wave components~\cite{Qi}, or in a spin-imbalanced Fermi gas with $s$-wave interactions~\cite{Bulgac}. In these cases, the $p$-wave pairing is either spin independent~\cite{Qi}, or driven by spin/density fluctuations and thus quite small in strength~\cite{Bulgac}. In our system, the induced $p$-wave pairing is due to the interplay of the $s$- and $p$-wave interactions, which gives rise to an appreciable pairing strength (see Fig.~\ref{fig3}(c1,c2)).

The hybridized nature of the S+P state is also reflected in the momentum-space density distributions $n(\cp k)=n_{\uparrow}(\cp k)+n_{\downarrow}(\cp k)$ and $\delta n(\cp k)=n_{\uparrow}(\cp k)-n_{\downarrow}(\cp k)$ in the $(k_{\perp},k_y)$ plane, with $k_{\perp}=\sqrt{k_z^2+k_x^2}$ (see Fig.~\ref{fig3}(e1,e2)). At $\cp k=0$, the $p$-wave pairing field vanishes, and the remaining $s$-wave pairing order requires $\delta n(0)=0$. The largest spin imbalance occurs at a finite momentum $k_{\perp}\sim k_u$, where the $p$-wave pairing, and through hybridization, the $s$-wave pairing as well, show the strongest amplitudes (see Fig.~\ref{fig3}(a,b1,b2)).

\section{Contacts}
Practically, the hybridized superfluid
can also be recognized in the measurement of contact~\cite{Tan}, a
physical quantity connecting the microscopic two-body physics with
the thermodynamics of a many-body system~\cite{Tan, Braaten, Zhang,
Yu, Ueda, Zhou}. Experimentally, the $s$- and $p$-wave contacts have been
successfully extracted in cold atoms experiments, respectively, in
an unpolarized spin-1/2 Fermi gas~\cite{Jin_contact1,Jin_contact2,
contact3}, and in a fully polarized Fermi gas~\cite{Toronto}.
In our system with mixed $s$- and $p$-wave interactions, the contactscan be individually defined through the adiabatic relations
under the grand canonical ensemble:
\begin{eqnarray}
\frac{\partial \Omega}{\partial (-1/a_s)} |_{\mu, h} &=&\frac{C_s}{4\pi M}, \label{Cs}\\
\frac{\partial \Omega}{\partial (-1/\nu_{\alpha})} |_{\mu, h} &=&\frac{C_{p,\nu}^{(\alpha)}}{4\pi M}, \label{Cv}\\
\frac{\partial \Omega}{\partial k_0} |_{\mu, h} &=&\frac{\sum_{\alpha}C_{p,R}^{(\alpha)}}{4\pi M} . \label{Cr}
\end{eqnarray}
Here, $C_s$ is the $s$-wave contact~\cite{Tan, Braaten,
Zhang}, and $C_{p,\nu}^{(\alpha)}, C_{p,R}^{(\alpha)}$ are
the $p$-wave contacts~\cite{Yu, Ueda,Zhou} respectively related to the scattering volume and the effective range. In the case of isotropic $p$-wave
interactions with $\nu_{\alpha}=\nu$, Eq.~(\ref{Cv}) can be replaced by
$\partial \Omega/\partial (-1/\nu) |_{\mu, h}
=C_{p,\nu}/4\pi M$, with $C_{p,\nu}=\sum_{\alpha}
C_{p,\nu}^{(\alpha)}$.

The contacts defined in Eqs.~(\ref{Cs}-\ref{Cr}) uniquely
determine the high-momentum tail of the spin-$\sigma$
($\sigma=\uparrow,\downarrow$) momentum distribution $n_{\sigma}(\cp k)$. Following the
standard derivations in Ref.~\cite{Tan, Braaten, Zhang, Yu, Ueda,Zhou}, we
obtain the following asymptotic behavior of $n_{\sigma}(\cp k)$ for $k_F\ll k\ll \Lambda$ ($k\equiv |{\cp k}|$, $k_F$ is the Fermi momentum):
\begin{align}
 n_{\downarrow}(\cp k) &\rightarrow
\frac{C_s}{k^4}, \label{nk_dn} \\
n_{\uparrow}(\cp k) &\rightarrow
\frac{C_s}{k^4} +\sum_{m} 4\pi |Y_{1m}(\hat{k})|^2 \Big[\frac{C_{p,\nu}^{(m)}}{k^2} + \frac{C_{p,R}^{(m)}}{k^4}\Big]. \label{nk_up}
\end{align}
Here, $\{ C_{p,\nu/R}^{(m)}\}$ $(m=0,\pm 1)$ are the projections of $\{ C_{p,\nu/R}^{(\alpha)}\}$ ($\alpha=x,y,z$) in the magnetic angular momentum space. For the experimentally relevant cases, we have $C_{p,\nu/R}^{(0)}=C_{p,\nu/R}^{(z)}$ and $C_{p,\nu/R}^{(\pm 1)}=C_{p,\nu/R}^{(x,y)}$. Eqs.~(\ref{nk_dn},\ref{nk_up}) show that the momentum distribution is highly asymmetric in spin. In particular, the distribution of the spin-$\uparrow$ component exhibits a quite non-trivial high-momentum tail, which is due to the involvement of both the $s$- and  the $p$-wave interactions. In contrast, the spin-$\downarrow$ component follows the same $1/k^4$ distribution as in the $s$-wave case, since the induced $p$-wave pairing within the spin-$\downarrow$ species is purely a low-energy effect and does not generate additional high-momentum or short-range physics. Given Eqs.~(\ref{nk_dn},\ref{nk_up}), both the $s$- and the $p$-wave contacts can be directly extracted from the time-of-flight measurement in experiments.

\begin{figure}[h]
\includegraphics[width=7.5cm]{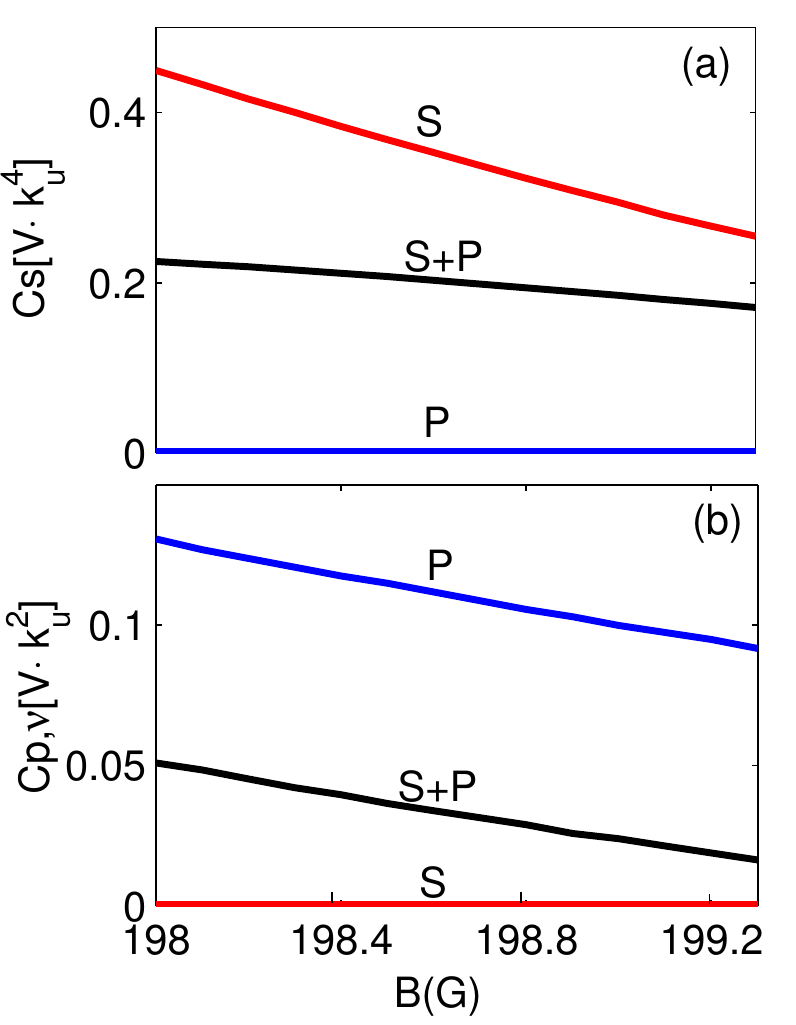}
\caption{(Color online). The $s$-wave (a) and the $p$-wave (b) contacts for the S, P and S+P states as the system crosses the $p$-wave resonance, with the magnetic field changing from $198$G to $199.3$G (assuming $\nu_{x,y}$ follow the variation of $\nu_z$ in Fig.~\ref{fig1} and $\nu_{\alpha}=\nu$). We choose $(\mu/E_u, h/E_u)=(0.3,-0.25), \ (0.25,0.25), {\rm \ and \ }(0.25,0)$ respectively for the S, P and S+P states.}
\label{fig4}
\end{figure}

In Fig.~\ref{fig4}, we show $C_s$ and $C_{p,\nu}$ across the $p$-wave FR, for the cases of pure $s$-wave superfluid (S), pure $p$-wave superfluid (P), and hybridized
superfluid (S+P). A distinguished feature of the
S+P hybridized superfluid is that both $C_s$ and $C_{p,\nu}$ are
finite, while we have $C_s=0$ for P and $C_{p,\nu}=0$ for S state.
Moreover, we have checked that for the S+P state, the $s$-wave contact
$C_s$ sensitively depends on the variation of the $p$-wave
scattering volume $(\nu)$ even with the $s$-wave scattering length ($a_s$) fixed. Similarly, the $p$-wave contact $C_{p,\nu}$ depends on
the variation of $a_s$ with $\nu$ fixed. All these properties of contacts
unambiguously confirm the co-existence and hybridization of the $s$- and the $p$-wave
pairings. Thus the contacts measurement can serve as another
effective tool for the detection of hybridized superfluid in
experiments.

\begin{figure}[h]
\includegraphics[width=8.5cm]{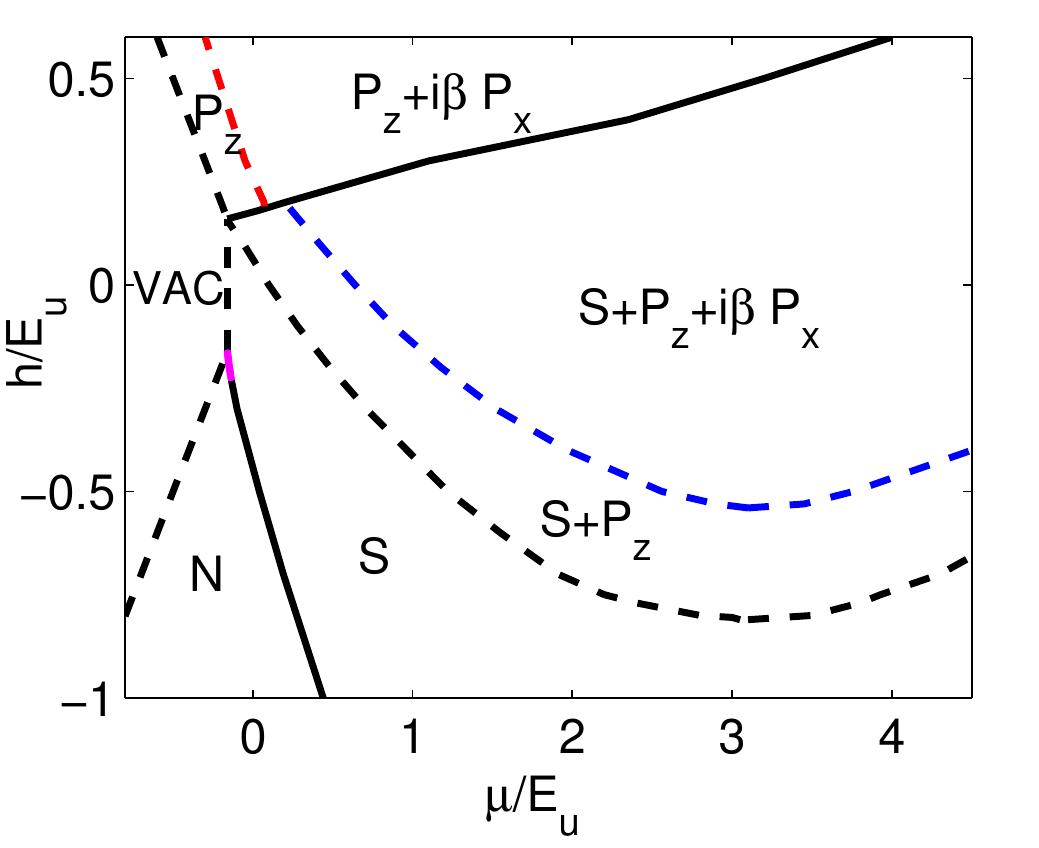}
\caption{(Color online). Ground-state phase diagram in the ($\mu,h$) plane for anisotropic $p$-wave interactions with $1/\nu_{z}=0$ and $1/\nu_x=1/\nu_y=-k_u^3$. The other parameters are the same as in Fig.~\ref{fig2}.  }
\label{fig5}
\end{figure}

\section{Effect of anisotropic p-wave interaction}
In the previous discussions, we have demonstrated the key features of the hybridized superfluid phase under an isotropic $p$-wave interaction. In $^{40}$K Fermi gases, the $p$-wave interactions are typically anisotropic (see Fig.~\ref{fig1}). In Fig.~\ref{fig5}, we map out the phase diagram in the $(\mu, h)$ plane under a typical anisotropic $p$-wave interaction: $1/\nu_z=0$ and $1/\nu_x=1/\nu_y=-k_u^3$. The only essential difference from the isotropic case is that the $p$-wave pairing states here (both the P and the S+P phases) have richer inner structures. Namely, the original $p\pm ip$ pairing (in Fig.~\ref{fig3}) is now replaced by $p_z$ ($u\neq 0, v=0$) or $p_z+i\beta p_x$ ($\beta <1, 0<v<u$) pairings, depending on the densities of two spin species. In this case, the hybridized superfluid phase (S+P) still occupies a considerable region in the phase diagram, and its key features, as discussed previously, are not qualitatively altered by the anisotropy of $p$-wave interactions.

\section{Summary and discussion}
To summarize, we have revealed a new type of fermion superfluid in spin-1/2 Fermi systems with hybridized $s$- and $p$-wave pairings. We demonstrate various non-trivial properties of such a hybridization, such as the phase locking of the induced $p$-wave pairing, the novel structure of pairing fields in momentum space, and the co-existence of $s$-wave and $p-$wave contacts. These features can be detected from the measurement of number distribution in real and momentum space in two-component $^{40}$K Fermi gases with very close $s$- and $p$-wave FRs near $B\approx 198$G.

All the features of the hybridized superfluid indicate that it is a highly entangled state that unifies different pairing symmetries in a single system. Such an entanglement emerges on the many-body level (rather than two-body), and is therefore a collective phenomenon due to Cooper pairs. Moreover, such a superfluid can accommodate large spin magnetization in a considerably broad parameter regime (see Fig.~\ref{fig1}), in contrast to previously discussed magnetized superfluids such as the Fulde-Ferrell-Larkin-Ovchinnikov state, whose stability region in three dimension is typically small on the phase diagram.

The discovery of the hybridized superfluid enriches our knowledge of pairing superfluidity in Fermi systems, and opens a promising avenue for achieving novel fermion superfluid with multiple partial-wave scatterings in cold atomic gases. For instance, in future studies, it is worthwhile to address the hybridized pairing in lower spatial dimensions, where the atom losses may be further reduced~\cite{Thywissen_talk} due to the much less singular two-body wave function at short range compared to the three dimensional case. Furthermore, as the $p$-wave interaction in lower dimensions can induce topological superfluid, the hybridization could be even more intriguing as it would involve not only different pairing symmetries, but also different topology classes.

\section*{Acknowledgements}
This work is supported by the National Natural Science Foundation of China (NSFC) (Grant Nos. 11374177, 11374283, 11421092, 11522545, 11534014), and the programs of Chinese Academy of Sciences. W. Y. acknowledges support from the ``Strategic Priority Research Program(B)'' of the Chinese Academy of Sciences, Grant No. XDB01030200.


\begin{thebibliography}{99}
\bibitem{note_zeromomenta} Here we have neglected the Fulde-Ferrell-Larkin-Ovchinnikov phase with finite total momentum of s-wave pairing, which occupies a very narrow parameter regime in the three dimension~\cite{FFLO, FFLO2, Sheehy, Hu}. As the $p$-wave pairing is between the same species of fermions, the zero-momentum pairing is expected to occupy the largest phase space among all pairings with different total momenta.
\end{thebibliography}
\end{document}